# Nanopore and nanoparticle formation with lipids, undergoing polymorphic phase transitions

Diana Cholakova, Desislava Glushkova, Slavka Tcholakova, Nikolai Denkov*

Department of Chemical and Pharmaceutical Engineering
Faculty of Chemistry and Pharmacy, Sofia University,
1 James Bourchier Avenue, 1164 Sofia, Bulgaria

*Corresponding authors:
Prof. Nikolai Denkov
Department of Chemical and Pharmaceutical Engineering
Sofia University
1 James Bourchier Ave.,
Sofia 1164
Bulgaria
E-mail: nd@lcpe.uni-sofia.bg
Tel: +359 2 8161639
Fax: +359 2 9625643

**ABSTRACT**

We describe several unexpected phenomena, caused by a solid-solid phase transition (gel-to-crystal) typical for all main classes of lipid substances – phospholipids, triglycerides, diglycerides, alkanes, *etc*. We discovered that this transition leads to spontaneous formation of a network of nanopores, spreading across the entire lipid structure. These nanopores are spontaneously impregnated (flooded) by water when appropriate surfactants are present, thus fracturing the lipid structure at a nano-scale. As a result, spontaneous disintegration of the lipid into nanoparticles or formation of double emulsions is observed, just by cooling and heating of an initial coarse lipid-in-water dispersion around the lipid melting temperature. The process of nanoparticle formation is effective even after incorporation of medical drugs of high load, up to 50 % in the lipid phase. The role of the main governing factors is clarified, the procedure is optimized, and the possibility for its scaling-up to industrially relevant amounts is demonstrated.

Solid lipid nanoparticles (SLN) are widely studied in recent years, because of their high potential for multiple applications, incl. controlled and localized drug release at specific cells, organs or tissues.[1-5] Nanoparticles are used also in foods, cosmetics and home care products to encapsulate, protect and deliver lipophilic components such as fragrances, flavors, vitamins and biologically-active lipids, *e.g.* prostaglandins, ω-3 fatty acids, and other unsaturated hydroxyl fatty acids.[6-9] The products containing nanoparticles have several advantages over those containing bigger particles. They ensure better bioavailability[10,11] in oral and parenteral delivery systems and better endocytosis uptake in targeted organs,[12] prolonged shelf-life due to better stability to particle aggregation and gravitational separation, and may be optically transparent which is important for many beverages and in some food and pharmaceutical applications.[1,6,7]

SLN are obtained from oil-in-water nano- (kinetically stable) and micro-(thermodynamically stable) emulsions. Two main types of preparation methods are distinguished, depending on the energy required for drop breakage: high-energy methods and low-energy methods.[1,6-8] In the high-energy methods, microfluidizers, high pressure homogenizers and ultrasonicators are used to obtain nanosized droplets. Most of the energy introduced in these homogenizers is lost as heat and sound. Only a very small fraction of the used energy, well below 0.01%, is used for the actual drop breakage process and the related surface energy increase.[13]

The low-energy methods include several phase-inversion methods.[1,6-8] Note that the most popular "phase inversion temperature" (PIT) method could be of high or low energy demand, depending on the amplitude of temperature variation needed. These later methods are system-specific and are applicable to limited number of systems. Furthermore, these methods usually require specific surfactants of rather high concentration, above ca. 10 wt %.[7]

Triacylglycerols (TAGs) are of outstanding interest in this area of research, as they are the main components in SLN for pharmaceutical, cosmetics and food applications.[1,3,8-10] On the other hand, the preparation of TAG nanoparticles is a particular challenge because TAG materials have relatively high interfacial tension and viscosity, at low polarity and water solubility.[7] Several examples for preparation of TAG nanoparticles were published in the literature, most of them based on the usage of two or more cycles of high pressure homogenization or sonication,[1,5,14-16] preparation of microemulsions upon heating and phase inversion.[5,17,18] Each of these methods has drawbacks and limitations when applied to TAGs, as explained above and in the original studies.

Searching for different methods to produce lipid nanoparticles, we unexpectedly discovered several related phenomena, all caused by a gel-to-crystal (α→β) lipid phase transition. We found that this transition leads to spontaneous formation of a network of nanopores, spreading across the entire lipid structure. When the lipid is in contact with aqueous surfactant solution, these nanopores are spontaneously impregnated by water when appropriate surfactants are present. As a result, the lipid structure bursts spontaneously into 20 to 100 nm nanoparticles, dispersed in the aqueous phase. This method of lipid nanoparticle formation is energy-efficient, scalable and can be applied to wide range of substances, incl. TAGs, phospholipids, diacylglycerols, and alkanes. The lipid could be loaded with actives of very high concentration (incl. drugs) without affecting the process. Thus we have created a convenient platform for studying lipid nanoparticles, incl. for drug research and formulation design.

**RESULTS AND DISCUSSION**

**Cold-burst phenomenon**

The cold-burst process developed in the current study produces nanoparticles, starting from micrometer sized oil emulsion droplets (*e.g.* of di- or triacylglycerols, phospholipids, alkanes), dispersed in aqueous solution of water-soluble and (possibly) oil-soluble surfactants with hydrophobic tails having number of carbon atoms $\geq C_{12}$. The initial coarse emulsions can be prepared by conventional emulsification methods, such as rotor-stator homogenization, mixer agitation, membrane emulsification or by simple hand shaking of the oil and aqueous surfactant solution closed in a container. In our experiments we prepared the initial coarse oil-in-water emulsions either by rotor-stator homogenization[19] or by membrane emulsification[20] when monodisperse drops were required. The specific initial drop size is not of particular importance and can be up to few hundreds of micrometers.

The method consists of two main steps. First, the coarse emulsion is cooled so that the dispersed drops freeze into solid lipid particles and, then, the dispersion with solid particles is either stored at temperature below the melting point of the lipid particles or heated up slowly to temperature around the lipid melting point. In both cases we observed that the lipid particles burst into multiple sub-micrometer and nanometer particles if appropriate surfactants ensuring low three-phase contact angle are used - see section Main factors below for explanation of the role of contact angle.

Let us illustrate the observed processes with the emulsion of $C_{12}$TG (trilaurin) droplets, dispersed in aqueous solution of the nonionic surfactants $C_{18}EO_{20}$ (1.5 wt. %) and $C_{18:1}EO_2$ (0.5 wt. %). When the aqueous dispersion containing frozen $C_{12}$TG particles is stored for 1 hour at 5°C, which is well below the bulk melting temperature[16] of $C_{12}$TG, $T_m \approx 46°C$, the micro-particles spontaneously burst into much smaller particles which literally "disappear" when observed under the optical microscope, because the formed nanoparticles are much smaller than the wavelength of the visible light, Figure 1a-d. This process of particle bursting was observed also when we heated the same $C_{12}TG$ dispersion with low heating rate, between ca. 0.5 and 2°C/min – see Figure 1e-h and Supplementary Movie 1.

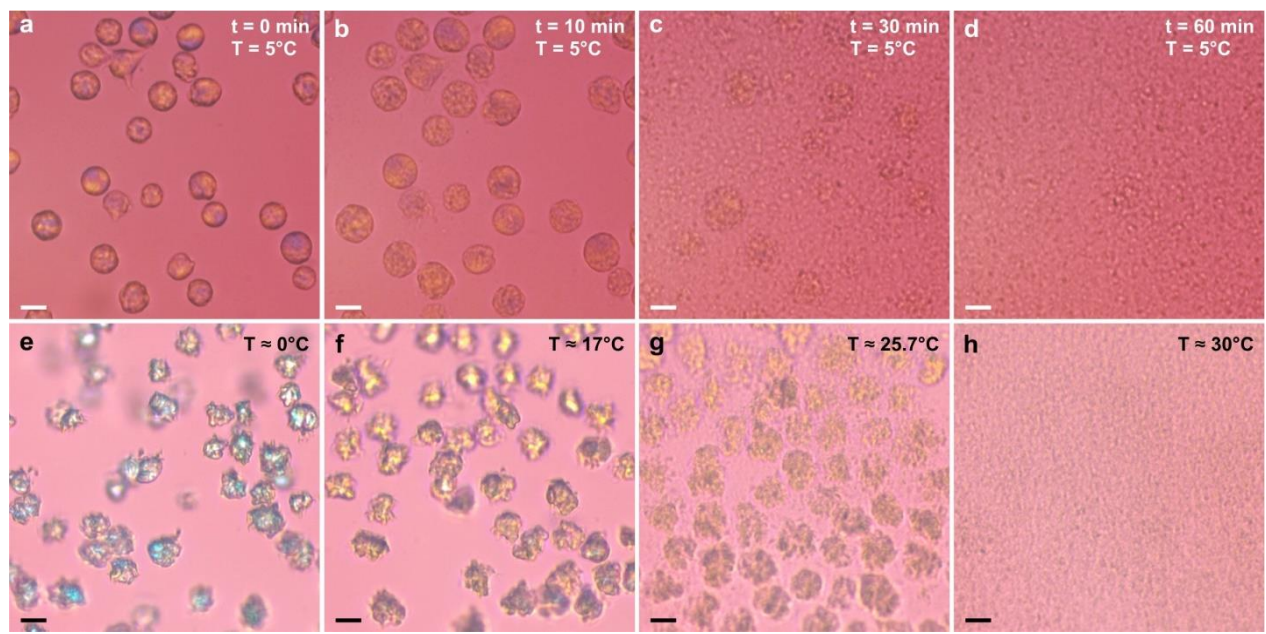

**Figure 1. Microscope images showing the spontaneous disintegration (cold-bursting) of frozen trilaurin ($C_{12}TG$) particles, dispersed in aqueous surfactant solution. a-d** $C_{12}TG$ particles stored for 60 min at $T$ = 5°C, far below the $C_{12}TG$ melting temperature of $T_m \approx 46°C$. The particles spontaneously disintegrate into much smaller particles with diameter < 0.4 μm. **e-h** $C_{12}TG$ particles observed upon heating at 2°C/min rate. The particles first increase their volume (f,g) and then disintegrate completely at $T \approx 30°C$. The aqueous medium contains 1.5 wt. % $C_{18}EO_{20}$ and 0.5 wt. % $C_{18:1}EO_2$ as nonionic surfactants. Scale bars, 10 μm.

As seen in Figure 1a,e, the initial frozen particles have bright colors when observed in polarized transmitted light, because of the alignment of the triglyceride molecules in the crystalline lattice. Upon storage at low temperatures, Figure 1b, or upon temperature increase (although still remaining below the particles' melting temperature), Figure 1f,g, the particles visually increase their size with time, while their colors almost disappear. The observed volume increase demonstrates that the aqueous phase penetrates into the interior of the frozen particles, see the explanations in section Mechanism below. Afterwards, each micrometer particle bursts into millions of smaller particulates, Figure 1d,h. As an additional example, Supplementary Movie 2 shows the spontaneous disintegration process of tripalmitin particles with initial diameter of $d_{ini} \approx$ 90 μm which burst into particulates with diameter of $\approx$ 0.4 μm after one freeze-thaw cycle only, *i.e.* from each initial 90 μm drop around 10 million individual small particles are formed at once, without any mechanical input into the system.

To clarify the capabilities of our approach for lipid nanoparticle formation, we studied more than 70 different combinations of triacylglycerols (TAGs) with chain lengths varied between 10 and 18 C-atoms and of other oils, including hexadecane, dilaurine and the phospholipid DPPC (see Supplementary Tables S1 and S2), emulsified in solutions of various nonionic and ionic surfactants (Supplementary Table S3). Briefly, the nonionic surfactants include: alcohol ethoxylates with linearly connected ethoxy units (chemical formula $C_nEO_m$, trade name Brij) with $n$ varied between 12 and 18 C-atoms and $m$ varied between 2 and 50; alcohol ethoxylated surfactants with sorbitan ring in the hydrophilic head ($C_nSorbEO_{20}$, Tween) with $n$ varied between 12 and 18. Most of the surfactants had saturated hydrophobic tails, but we tested also surfactants with $C_{18}$ unsaturated tails with double bond in the middle (denoted as $C_{18:1}$). Two monoglyceride-diglyceride mixtures (MG-DG, denoted as SM1 and SM2) and $C_{18}MG$ were also tested as surfactants. Two common ionic surfactants were also studied – the anionic SDS (sodium dodecyl sulfate) and the cationic CTAB (cetyl trimethyl ammonium bromide) to check for the role of the surfactant head-group charge.

The selection of the above substances and their combinations was based on the following reasoning. The initial surfactant screening was made with $C_{14}TAG$ (trimyristin), because this TAG has very convenient phase transition temperatures for the experiments of interest: $T_{m,\beta} \approx 56\text{-}57°C$ and $T_{m,\alpha} \approx 33°C$. Experiments with shorter and longer chain TAGs were also made ($C_{10}$ to $C_{18}$) to demonstrate the wider applicability of the method. To prove that the method works with TAG mixtures, we mixed two or several TAGs with number of C-atoms similar to those found in the plant oils and animal fats.

The initial surfactant screening was made with a wide range of water-soluble surfactants having chain length similar (+/- several carbon atoms) to that of $C_{14}TAG$, because these surfactants were expected to be the most compatible with the oil, *e.g.* in forming adsorption layers with intercalated TAG molecules. These initial experiments showed that various surfactants, with different ionic and nonionic head-groups, induce the cold-burst process, though with diverse efficiency. After this initial screening of water-soluble surfactants, we tested whether the method efficiency is improved when oil-soluble surfactants are added. Indeed, we found that the combinations of water-soluble and oil-soluble surfactants is often more efficient than the individual surfactants (synergy) and made additional series of experiments to optimize several surfactant mixtures of this type. All components in the tested systems were selected to be permitted in various potential applications: drug delivery, cosmetics, personal care, foods and/or beverages. Summary for the behavior of various triglyceride-surfactant combinations, studied as described above, is presented in Table 1 and in Supplementary Table S4.

**Table 1.** Representative results about the efficiency of cold-burst process with $C_{14}TG$ oil drops (initial diameter between ca. 5 µm and 20 µm), dispersed in various surfactant solutions. Legend used in this table: the sign "-" means that the cold-burst process is not observed; the number of pluses increases with the efficiency of the cold-burst process; one "+" means that some disintegration is observed but it is note very efficient and many micrometer drops remain in the sample; "+++" denotes very efficient cold-burst process in which all initial drops disintegrated to submicrometer entities after one cooling and heating cycle; "++" denotes intermediate behavior; "W/O/W" means that cold-bursting is not observed, instead, double water-in-oil-in-water emulsion drops are formed.

| Oil-soluble surfactants, wt. % | Water-soluble surfactants, concentration in wt. % | | | |
|---|---|---|---|---|
| | **no** | **0.5% SDS** | **1.5% $C_{12}SorbEO_{20}$** | **1.5% $C_{18}EO_{20}$** |
| **no** | - | W/O/W | + | ++ |
| **0.1% $C_{12}EO_4$** | - | + | ++ | ++ |
| **0.5% $C_{12}EO_4$** | - | +++ | +++ | +++ |
| **1% $C_{18:1}EO_2$** | - | +++ | +++ | +++ |
| **0.5% SM1** | - | +++ | +++ | +++ |

To check whether this cold-burst process can be scaled-up to bulk TAG dispersions and to see what the minimal achievable size is of the formed nanoparticles, we performed experiments with batch dispersions contained in glass bottles, Figure 2. Several TAG-surfactant combinations were tested, as described in Figure 2. After one cooling and heating cycle, we measured by dynamic light scattering the volume-averaged particle diameter to be 410 ± 35 nm and the number-averaged diameter to be 300 ± 55 nm for all these systems, with oil weight fraction varied between 0.1 and 20 wt. %, Figure 2. After two cycles, the particle size decreased further down to 170 ± 65 nm by volume. After several freeze-thaw cycles, the minimal particle diameter obtained was between 20 and 200 nm, depending on the specific TAG-surfactant combination, Figure 2b. As expected, the samples containing ≈ 20 nm particles were completely transparent and those with ≈ 100 nm were translucent, although the initial dispersions containing micrometer particles were milky white, Figure 2c,d.

Note that the dispersions prepared by the above procedure are kinetically stabilized – they do not form spontaneously, if the TAG and the surfactant solution are mixed and stored for a long period at temperature above the TAG melting temperature. In other words, the phase transition energy of the particle freezing is used in our procedure to create nanoparticles with excessive surface energy, proportional to the inverse particle radius.

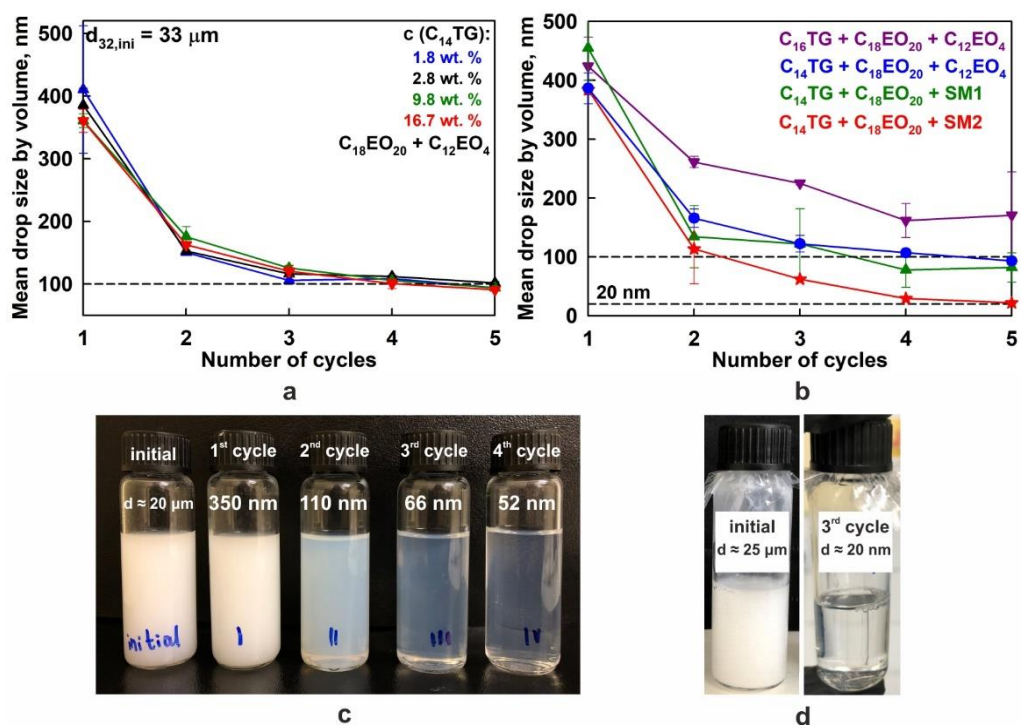

**Figure 2. Drop diameter in bulk TAG dispersions subject to particle freeze-thaw (FT) cycles.**
**a** Mean drop diameter by volume, as a function of the FT cycle number, for $C_{14}TG$ dispersions with different particle concentrations. The initial drops with diameter $d_{ini} \approx 33$ μm were dispersed in aqueous solution of the nonionic surfactants $C_{18}EO_{20}$ and $C_{12}EO_4$ (3:1 w/w). The blue, black and green points (up to 9.8 wt % oil) are obtained at 4 wt. % total surfactant concentration, the red points for 16.7 wt % oil – at 6 wt. %. **b** Mean drop diameter by volume for several TAG-surfactant combinations: $C_{16}TG$ with $d_{ini} \approx 10$ μm (purple) and $C_{14}TG$ with $d_{ini} \approx 33$ μm (blue) dispersed in $C_{18}EO_{20} + C_{12}EO_4$ solution (3:1 w/w); $C_{14}TG$ with $d_{ini} \approx 33$ μm (green) dispersed in surfactant solution of $C_{18}EO_{20}$ + SM1; $C_{14}TG$ with $d_{ini} \approx 20$ μm (red) dispersed in surfactant solution of $C_{18}EO_{20}$ + SM2. Each point is average over ≥ 3 independent experiments. **c** Picture of bottles containing $C_{14}TG$ particles in $C_{18}EO_{20} + C_{18}MG$ solution, as observed initially and after 1 to 4 FT cycles. After the 3rd cycle, the mean drop diameter is ≈ 66 nm. **d** $C_{14}TG$ dispersion in $C_{18}EO_{20} + C_{16}EO_2$ surfactant solution: initial sample with $d_{ini} \approx 25$ μm (left) and after 3 FT cycles with particle diameter ≈ 20 nm (right).

**Mechanism**

To explain the mechanism of the observed process of particle disintegration we start with the observation that the liquid drops of TAGs cream toward the top of their emulsions, because they have lower mass density than that of the continuous water phase. However, after storing the dispersions of frozen TAG particles for a certain period (minutes to hours), we observed that these particles sank at the bottom of the container, *i.e.* the TAG solid particles increased their mass density upon storage which became higher than that of water, see Supplementary Movie 3. Similar process of particle sinking was observed also during the observations with optical microscopy, see Supplementary Movie 4. This mass density increase after certain period of storage at low temperature is a direct indication that the molecular packing inside the frozen TAG particles changes with time, as a result of solid-state phase transition in the particle interior.

Previous structural studies revealed that TAGs form mainly three polymorphic solid phases: α-phase which is the least stable and most disordered one, with the lowest melting temperature; β-phase which is the most ordered one; and β'-phase which has intermediate properties.[16,21-23] The mass density of β-phase of $C_{14}TG$ is 1.05 g/ml, while the density of the isotropic liquid phase is 0.8722 g/ml.[21] Therefore, the observed density change is certainly due to a solid-state α → β (or α → β') polymorphic phase transition inside the TAG particles, as confirmed by SAXS and WAXS measurements, Figure 3a,b. Upon such transition, the TAG molecules pack better, thus increasing the particle mass density above that of water.

The polymorphic transitions α → β (or α → β') can occur in two different ways:[23] if the sample is stored for a long period at temperatures below the melting temperature of the α-phase, then a solid-solid (SS) state transition, $\alpha \xrightarrow{SS} \beta$ (or β'), may occur. Alternatively, if the temperature of the sample is increased above the melting temperature of the α-phase, but still kept below the melting temperature of the β (or β') phase, then the so-called "melt-mediated (MM) transition" $\alpha \xrightarrow{MM} \beta$ (or β') occurs.[23] These two options explain why we could observe particle disintegration while using either of the two protocols – at fixed temperature below the melting point of the lipid or upon slow heating up to melting of the lipid particles.

To explain the mechanism of particle bursting, we refer to several studies which reported the formation of nano-voids at the grain boundaries between the crystal domains in the frozen TAG phases.[24-27] The formation of these nano-voids causes the so-called "negative pressure effect", resulting from the local contraction of the nano-crystallites in the course of the solid-state polymorphic transitions.[25] In other words, the polymorph transitions lead to formation of a nano-porous internal structure in the lipid phase, containing nano-sized crystalline β-domains which are separated by nano-voids.

Combining this structural information with our observations, we reveal the following mechanism of spontaneous nanoparticle formation in the cold-burst method, Figure 3c. Upon rapid freezing, the triglyceride droplets crystallize first in the less stable α-form, because this is the phase with the lowest nucleation energy. When the sample is stored at low temperature for a certain period or upon slow heating, α → β polymorphic transition takes place, possibly passing through the intermediate β'-phase. Thus, the TAG molecules rearrange into more ordered crystalline β-domains, with nano-voids separating them. These nano-voids form a continuous 3D porous network inside the lipid particles and have a "negative pressure" which sucks aqueous phase into the particle interior – the surfactant solution fills the nano-porous structure between the crystalline lipid domains of the frozen particles and creates repulsion between these domains. We observe this process in the microscope as fading of the particle color, combined with an increase of the particle volume. When appropriate surfactants ensuring low contact angles are used, the water penetration leads to complete separation of the nano-crystalline domains into numerous small individual particles, Figure 3c (upper series of the final two schematics).

Alternatively, if the surfactant is unable to stabilize the crystalline domains against their coalescence in the moment of lipid melting, the water which has penetrated into the lipid porous structure remains captured (trapped) in the form of water droplets inside the bigger oily drop, thus forming water-in-oil-in-water (W/O/W) double emulsion, see Figure 3c (lower series of the final two images) and Supplementary Movie 5.

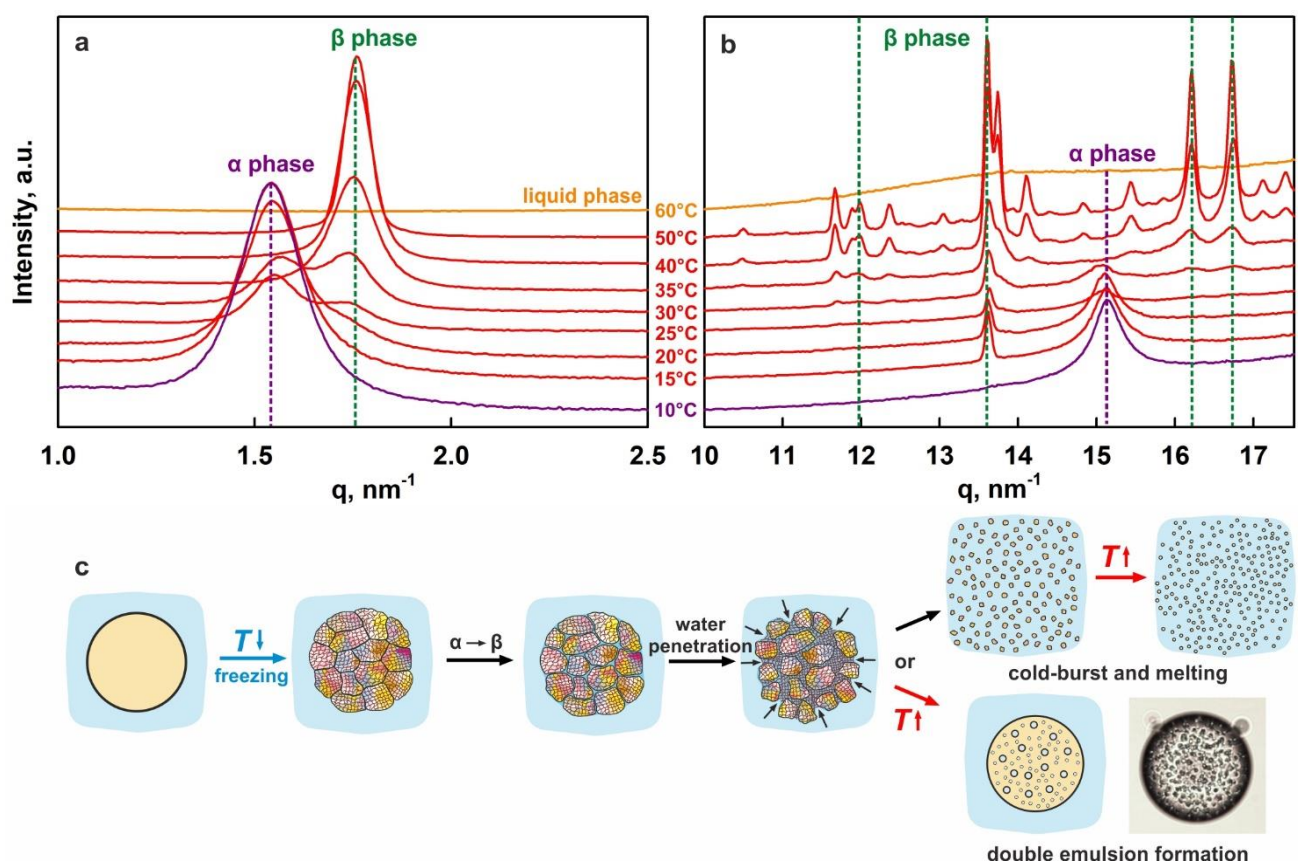

**Figure 3. Mechanisms of particle bursting and double emulsion formation. a-b** SAXS (a) and WAXS (b) signals obtained upon heating of $C_{14}TG$ dispersion which has been crystallized rapidly by inserting a pre-heated sample at 60°C into a cool chamber with $T = 10$°C. Only α-phase is present in this sample initially. Upon heating, it quickly transforms into more stable β-phase. Curves are shifted with respect to *y*-axis for clarity. The aqueous phase contains 1.5 wt. % $C_{18}EO_{20}$ + 0.5 wt. % $C_{12}EO_4$ as nonionic surfactants, heating rate 2°C/min. **c** Schematics of the mechanisms for formation of nanopores, nanoparticles and double emulsion drops, see text for further explanations.

**Main factors**

Based on the mechanism described above, we explain now the role of the main factors affecting the observed particle bursting and the strategies for its optimization.

*(a) Role of surfactant type and three-phase contact angle*

We observed under the microscope that the aqueous phase always penetrated into the frozen particles upon α → β phase transition, no matter what the surfactant molecular structure was and in which phase (oil and/or water) the surfactants were initially dissolved or dispersed. However, depending on the specific oil-surfactant combination, the particle dispersions behaved differently. Most efficient particle disintegration was observed when both oil-soluble and water-soluble surfactants were both present in the aqueous phase. The oil-soluble surfactant could be pre-dispersed as surfactant particles with limited solubility in the aqueous phase or could be completely solubilized (incorporated) inside the micelles of the water-soluble surfactant. Alternatively, the oil-soluble surfactant could be dissolved in the oily phase.

Note that the commercially available nonionic alcohol ethoxylated surfactants (*e.g.* of Tween, Brij, Span or Lutensol series) are mixtures of molecules with different number of ethoxy units, *i.e.* they usually contain certain fractions of oil-soluble and water-soluble components. Therefore, if water-soluble surfactant is used at sufficiently high concentration, no separate oil-soluble surfactant is needed – the latter is already present in the commercial surfactant. Similarly, if oil-soluble surfactant is used in high concentrations, it can contain water-soluble fraction which could lead to cold-bursting – see Figure 4e-g for example.

Our experiments showed that the key physicochemical parameter which determines whether the particle disintegrates into nanoparticles or, alternatively, double emulsion drops are formed, Figure 3c, is the ability of the aqueous surfactant solution to wet the crystalline domains of the frozen lipid and to stabilize them against coalescence upon lipid melting. To quantify this property, we measured the three-phase contact angles, θ, for drops of aqueous surfactant solutions, placed on top of a solid trimyristin ($C_{14}TG$) layer which mimicked the frozen lipid domains, see Methods section.

The contact angle for drops of pure water, placed on frozen $C_{14}TG$ substrate, was θ ≈ 109 ± 4° for $C_{14}TG$ crystallized in α-phase and θ ≈ 99 ± 7° for $C_{14}TG$ crystallized in β-phase, in agreement with the literature data, obtained by a similar method.[28] Such high values, θ > 90°, reflect the hydrophobic nature of the lipid $C_{14}TG$ surface. In the presence of surfactants, the difference between the contact angles, measured on α- and β-phases of $C_{14}TG$, is even bigger, while preserving the trend that the contact angle on the β-phase is always smaller as compared to that on the respective α-phase, Figure 4a. In other words, α → β transition makes the lipid crystals more hydrophilic.

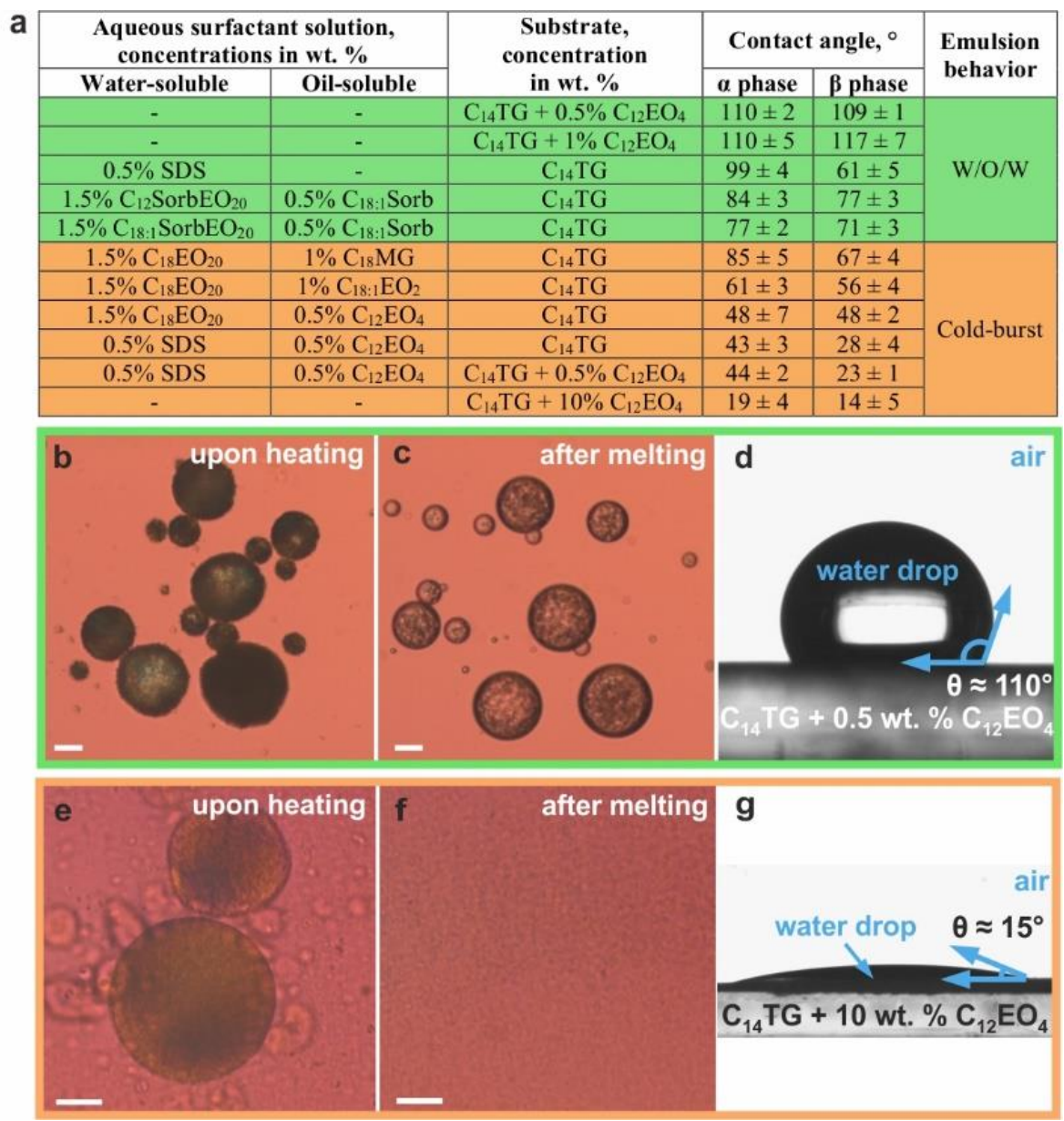

Figure 4. a. Three-phase contact angles of drops of surfactant solutions on solid $C_{14}TG$ substrate in air for various surfactant systems. Systems marked in green have large contact angle and form double W/O/W emulsion drops. Systems marked in orange have small contact angle and exhibit intensive cold-burst process. b,c. Microscope images of $C_{14}TG$ (+1 wt.% $C_{12}EO_4$) lipid particles, dispersed in pure water. Upon heating, water penetrates inside the particles as evidenced by the change in particle appearance – the initial particles were colorful but they became dark as seen in (b). However, this surfactant is unable to stabilize the melting lipid domains and, therefore, water-in-oil-in-water (W/O/W) emulsion drops are formed after melting. d. Contact angle θ ≈ 110° for this system. e,f. Images of $C_{14}TG$ + 10 wt. % $C_{12}EO_4$ particles, dispersed in pure water. Upon heating, water penetrates inside the particles. At such high concentration, the surfactant stabilizes well the melting lipid domains against coalescence and, therefore, complete particle disintegration is observed upon lipid melting. g. The contact angle is θ ≈ 15° for this system. Scale bars, 20 μm.

Such contact angle measurements allowed us to clarify that intensive disintegration process is typically observed for systems in which $\theta \lesssim 50°$ for substrates in β-phase. Examples for such systems are $C_{14}TG$ particles dispersed in aqueous solution containing 1.5 wt. % $C_{18}EO_{20}$ + 0.5 wt. % $C_{12}EO_4$; $C_{14}TG$ in 1.5 wt. % $C_{18}EO_{20}$ + 1 wt. % $C_{18:1}EO_2$; and $C_{14}TG$ in 0.5 wt. % $C_{12}SorbEO_{20}$ + 0.5 wt. % $C_{12}Sorb$.

In contrast, for the systems with $\theta \gtrsim 100°$, double emulsion drops are typically formed when the oil-soluble surfactant was present in the oily phase. Example for such system are the $C_{14}TG$ particles, containing pre-dissolved surfactant $C_{12}EO_4$ (introduced in the $C_{14}TG$ phase before preparing the lipid dispersion) when its concentration is below ca. 5 wt. %.

Interestingly, varying only the concentration of $C_{12}EO_4$ in the $C_{14}TG$ particles, we could switch the system behavior between the above two extremes. Double emulsion drops were formed at low $C_{12}EO_4$ concentrations in the oil phase, Figure 4b,c. When $C_{12}EO_4$ concentration in the oil phase was increased up to 10 wt. %, intensive drop disintegration was observed, Figure 4e,f. This switch in behavior is also explained with changes in the three-phase contact angle: for water drops placed on $C_{14}TG$ + 0.5 wt. % $C_{12}EO_4$ substrate $\theta \approx 110°$, whereas $\theta \approx 15°$ for $C_{14}TG$ + 10 wt. % $C_{12}EO_4$ substrate, Figure 4d,g, thus confirming that the contact angle is the key factor.

For intermediate contact angles, $60° \lesssim \theta \lesssim 80°$ (on β-substrate), the different systems varied in behavior, because other factors started to interfere. For example, in some of the systems we observed the cold-burst process for the outer layer of the frozen lipid particle, but it did not lead to complete disintegration of the biggest particles – some micrometer sized drops remained after the first cooling-heating cycle. In this case, the used protocol did not allow the surfactant solution in these systems to penetrate sufficiently deep into the particle interior and to stabilize the newly formed small lipid particles against coalescence. In other systems falling in the same range of intermediate contact angles, we observed the formation of double emulsion droplets, but the number of the trapped water droplets was significantly lower while their size was bigger, as compared to the reference systems with large contact angles (*e.g.* $C_{14}TG$ + 0.5 wt. % $C_{12}EO_4$ with $\theta \approx 110°$).

It would be very useful to reveal the effects of the chain-lengths of the TAGs and of the surfactants inducing cold-burst phenomenon or double emulsion formation. The TAGs with longer saturated chains have higher melting temperatures and higher temperature of α→β polymorphic phase transitions.[21] Therefore, one can tune the temperatures at which the phenomena of interest would occur by selecting an individual TAG or TAG mixtures with appropriate chain lengths. The role of the surfactant chain-length in the studied phenomena is less clear. We performed series of experiments in which we varied the chain-length of the water-soluble surfactants in mixtures with given oil-soluble surfactant and *vice versa*. However, no clear trend was observed about the effect of surfactant chain-length on the efficiency of the cold-burst method, see Supplementary Table S4.

*(b) Role of cooling-heating protocol*

Another important factor for intensifying particle disintegration is the cooling protocol which affects the size and type of the crystalline domains formed upon oil freezing. When we applied rapid cooling (ca. 15°C/s) followed by slow heating (rate < 2°C/min), we observed very efficient particle disintegration for the system $C_{14}TG$ dispersed in solution of 1.5 wt. % $C_{18}EO_{20}$ + 0.5 wt. % $C_{12}EO_4$, because many smaller crystal domains were formed at high cooling rates.[29] Upon slower cooling, fewer and larger in size crystal domains are formed and, respectively, bigger nanoparticles are obtained after heating. These trends confirm that the size of the initially formed α-domains is reflected into the subsequently formed β-domains upon the polymorphic transition and into the size of the final nanoparticles.

In the other extreme, experiments with pre-formed micrometer-sized monocrystals of $C_{14}TG$ β-phase, dispersed in the same surfactant solution, 1.5 wt. % $C_{18}EO_{20}$ + 0.5 wt. % $C_{12}EO_4$, showed no disintegration into smaller particles, Figure 5, because the β-phase is thermodynamically stable upon temperature variations below the lipid melting point.

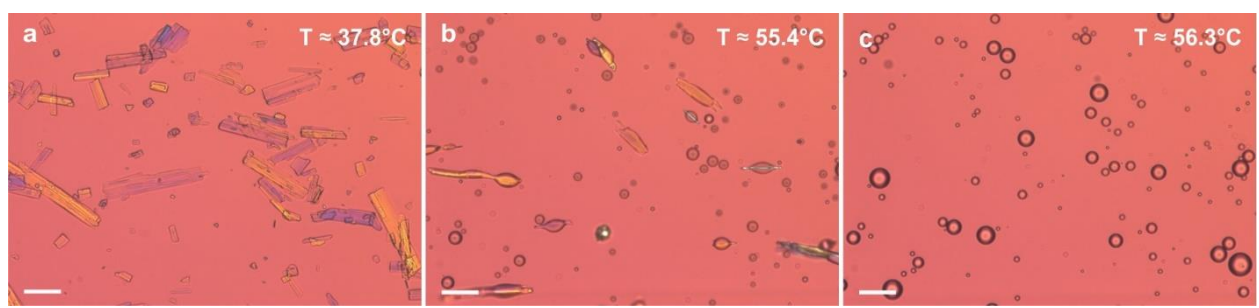

**Figure 5. Melting of $C_{14}TG$ monocrystals in β-phase without disintegration.** This sample was prepared by directly dispersing $C_{14}TG$ micro-particles, as received from the producer (without pre-melting) into the surfactant solution of 1.5 wt. % $C_{18}EO_{20}$ + 0.5 wt. % $C_{12}EO_4$. The crystallites fell directly onto the container bottom, *i.e.* they were in β-phase. The microscopy images are taken during heating of such dispersion, placed in a glass capillary. **a.** Microcrystals at temperature well below $T_m$. **b.** Crystal melting. **c.** Emulsion obtained after particle melting – single drop was formed from each individual crystal. No crystal disintegration was observed, because no α → β solid-phase transition occurred in this system. Scale bars, 50 μm.

The protocol of heating also affects the efficiency of particle disintegration, especially for larger in size initial particles. When the phase transition occurs entirely in the solid-state without intermediate melting, $\alpha \xrightarrow{SS} \beta$, the aqueous phase has more time to penetrate in between the crystalline domains. Thus, lower heating rates provide longer penetration time for the aqueous phase and lead to more efficient particle disintegration. In contrast, the melt-mediated transition which occurs at higher heating rate, $\alpha \xrightarrow{MM} \beta$, gives much shorter time for water penetration and may result in larger final particles.

Most efficient disintegration was observed in a specially designed protocol, in which the frozen lipid particles were slowly heated up to a given temperature of storage, $T_{st} < T_m$, and then stored at this temperature for a period of ca. several hours. When the latter protocol for nanoparticle formation is used, the specific value of $T_{st}$ controls the rate of particle bursting, Figure 6b-g. Also, longer storage times at $T_{st}$ are needed for the bigger particles, due to the larger penetration "depths" which should be overcome by the aqueous surfactant solution for such particles. DSC measurements confirmed that the particle disintegration observed at $T_{st} < T_m$ does not include any crystal melting, viz. the particles disintegrate while being entirely in a solid state – the same melting enthalpy was measured in experiments with and without storing the samples at temperature $T_{st}$, Figure 6a.

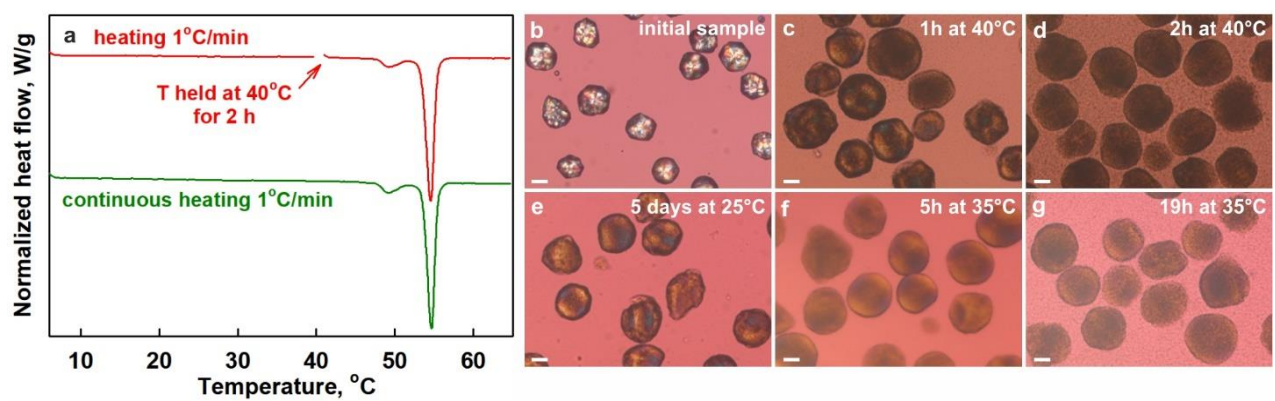

**Figure 6. Particle disintegration without melting. a** DSC curves of two identical samples heated using two different protocols. The green curve presents the signal obtained upon continuous heating at 1°C/min, while the red curve – continuous heating to 40°C, holding the temperature constant for 2 h, followed by a second heating period. The melting peaks are identical for the two samples. Curves are shifted with respect to *y*-axis for clarity. **b-g** Microscopy images of samples held at different constant temperatures for various periods of times, as indicated on the images. **b** Image of the initial frozen particles. **c,d** Images obtained upon holding the temperature at 40°C for 1 and 2 h. The temperature protocol reproduces the one used for the DSC measurements in **a**. **e** Water penetration after 5 days storage at 25°C is almost identical to the one observed in (c). **f,g** Upon storage at 35°C, drop disintegration is observed within several hours. The tested system is: $C_{14}TG$ particles with $d_{ini} \approx 33$ μm, dispersed in surfactant solution of 1.5 wt. % $C_{18}EO_{20}$ + 0.5 wt. % SM2. Scale bars, 20 μm.

*(c) Control of particle size in the cold-burst method*

From the experimental results presented above we could conclude that the particle size can be tuned in a relatively wide range by varying: (1) the number of cooling-heating cycles applied to the specific sample; (2) surfactant-to-oil ratio, because certain amount of surfactant is needed to cover the increased surface of the formed smaller particles with dense surfactant adsorption layer; and (3) the specific surfactant(s) used for a given oil. As illustrative examples of these trends, one can use the experimental data shown in Figure 2a,b which illustrate the effects of these factors: particles with diameter around 400 nm are obtained usually after one cycle, while smaller particles are obtained after the 2nd and 3rd cycles, depending on the specific oil-surfactant combination studied and on the surfactant concentration used.

The amount of water-soluble surfactant needed to cover the nanoparticle surface could be estimated approximately by adapting an approach, developed years ago for description of the drop size in the so-called "surfactant-poor regime" in turbulent emulsification.[30,31] This approach provides the following explicit equation, relating the particle size with the surfactant concentration and oil volume fraction, in the systems for which the particle size is limited by the available surfactant:

$$d_{32} \approx \frac{6\Phi}{(1-\Phi)} \frac{\Gamma_M}{C_S} \qquad (1)$$

Here $d_{32}$ is the mean surface-volume diameter, $\Phi$ is the particle volume fraction, $C_S$ is the initial concentration of the water-soluble surfactant in the aqueous phase, and $\Gamma_M$ is the surfactant adsorption in the dense adsorption monolayer which is typically $\approx$ 2-3 mg/m$^2$ for low-molecular mass surfactants like those used in the current study. Taking $\Gamma_M \approx$ 3 mg/m$^2$ as a reasonable value for our systems, we estimate that $C_S \geq$ 20 kg/m$^3$ = 2 wt. % is needed for particles with $d_{32} \approx$ 100 nm and $\Phi \approx$ 10 %. These estimates are in reasonable agreement with our experimental results in which we had to use a surfactant concentration in the range between 1 and 6 wt. % to obtain particles in the range of 100 nm or smaller.

The contribution of the oil-soluble surfactant in the particle size is more difficult to assess, as these surfactants could participate in mixed adsorption layers, thus reducing the needed concentration of water-soluble surfactant to form a dense adsorption layer. However, big fraction of the oil-soluble surfactant could remain dissolved in the oil drop interior without participating in the mixed adsorption layer. Furthermore, if the oil-soluble surfactant has higher surface activity, it could displace the water-soluble surfactant from the adsorption layer and act as demulsifier, instead of assisting particle stabilization.

**Method versatility and applicability**

Our experiments revealed that the process of nanoparticle formation, described above, is applicable to wide range of TAG-surfactant combinations, see Tables 1, S4 and S5. In this way, we prepared nanoparticle suspensions with all monoacid TAGs with chain length varied between 10 and 18 carbon atoms. Also, increasing the temperature above the TAG melting point allowed us to obtain nanoemulsions.

We tested successfully various TAG-surfactant combinations which are (potentially) applicable in different industries. For example, TAG particles, stabilized by $C_{12}SorbEO_{20}$ and mono- and/or diglycerides (surfactant mixture SM1 or SM2), are applicable in food and beverage products. Particles, stabilized by $C_{18}EO_{20}$ and $C_{12}EO_4$, can be used as delivery vehicles for tumor targeting drug delivery, as these surfactants are known to overcome the multidrug resistance in cancer.[32-34] SLN of 20-30 nm, similar to those obtained in the current study, were shown to have longest blood circulation half-life and higher tumor accumulation.[17,35]

We performed also experiments with TAG mixtures to show that the cold-burst process is not limited to individual substances but it is also observed with mixtures. TAGs with chain lengths between 12 and 16 C-atoms were chosen as these are the triglycerides commonly found in plant oils and animal fats. For example, we observed spontaneous particle bursting for the binary mixture $C_{12}TG + C_{14}TG$ (1:1 w/w) dispersed in aqueous solution of 1.5 wt. % $C_{18}EO_{20}$ + 1 wt. % SM2, as well as in various other surfactants, see Supplementary Movie 6 and Supplementary Table S5. The particles of the triple mixture $C_{12}TG + C_{14}TG + C_{16}TG$ (1:1:1 w/w/w), dispersed in aqueous solution of 1.5 wt. % $C_{18}EO_{20}$ + 0.5 wt. % SM1, also burst efficiently. These latter results are important from the viewpoint of practical applications, because most industrially relevant substances are mixtures of lipid molecules with different chain lengths. In particular, TAG mixtures are preferred for SLN preparation in the drug delivery systems, because polymorph transitions in pure TAGs may cause expulsion of the incorporated drug molecules, thus leading to drug precipitation in the aqueous phase and significant decrease of the drug load.[5,6]

To test the applicability of the cold-bursting method to other classes of lipid substances, we performed similar experiments with aqueous dispersions of α,α'-dilauryl glyceride ($C_{12}DG$, diglyceride), dipalmitoyl phosphatidylcholine (DPPC, phospholipid) and hexadecane ($C_{16}$ alkane). The tested diglyceride and phospholipid systems showed very similar behavior to that observed with the TAG systems, see Supplementary Movie 7. We obtained nanoparticles even after one cooling-heating cycle which showed that the method is very efficient for diglycerides and phospholipids. With alkanes, similar process of particle disintegration was observed but a larger number of cooling-heating cycles was needed to obtain small nanoparticles.

To check whether this procedure can be used for obtaining nanoparticles loaded with bio-active molecules, we performed experiments with TAGs in which model medical drugs (progesterone or fenofibrate) were pre-dissolved in high concentrations. We observed that the process of particle disintegration remained unaffected for up to ≈ 30 wt. % of drug content – the efficiency of the process and the lipid particle size were the same when compared to the drug-free TAGs, see Figure 7. At 50 wt.% drug loading, the particle bursting was still rather intensive but its efficacy was somewhat lower – along with the prevailing nanoparticles we observed a small fraction of residual micrometer sized particles after the first cooling-heating cycle.

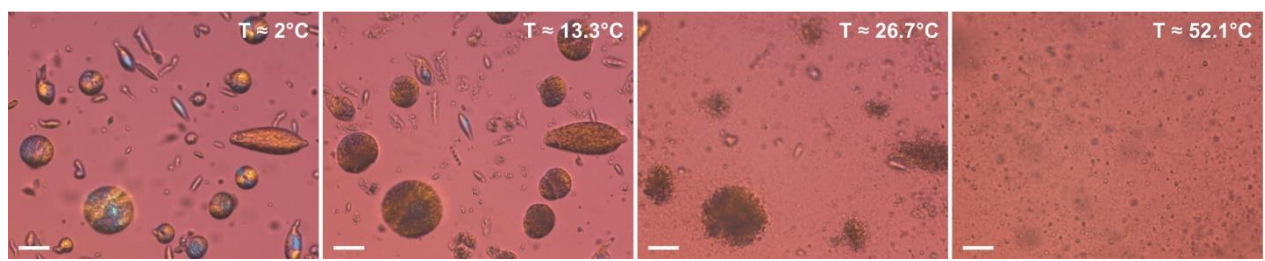

**Figure 7. Cold-burst process observed with trimyristin particles, containing 30 wt. % fenofibrate drug.** Efficient cold-burst process is observed in 1.5 wt. % $C_{12}EO_{23}$ + 0.5 wt. % $C_{12}EO_4$ surfactant solution. Water penetration begins at ca. 10°C and the smaller initial particles are completely disintegrated at $T \approx 26°C$. After melting, the main fraction of the drops is of submicrometer size. Experiment performed at 1°C/min heating rate. Scale bars, 20 μm.

**CONCLUSIONS**

In conclusion, we discovered simple method for preparation of solid lipid nanoparticles and nanoemulsion droplets. Particles and drops with mean volume diameter in the range between 20 and 200 nm can be prepared on-demand, using appropriate surfactants ensuring low contact angles, and one to several cooling-heating cycles below/around the melting temperature of the lipid. The mean size of the particles can be controlled in wide range by the selection of surfactants and their concentration, and by the number of cooling-heating cycles employed. Alternatively, one can select surfactants with high contact angles which lead to the formation of double W/O/W emulsion. We show that both processes are resulting from the formation of a 3D network of nanopores in the lipid structure, triggered by α → β polymorphic phase transition which is typical for many lipid substances. We show that, indeed, this approach can be applied to different classes of lipid substances and their mixtures. The method can be used for preparation of SLN, loaded with useful actives of very high concentration, such as medical drugs, magnetic nanoparticles, fluorescent molecules, *etc.*, which are all of highest interest for theranostic applications. The method can be used to generate lipid particles of direct interest to the pharmaceutical, cosmetic, food and beverage industries.

## METHODS

**Materials**

The producers and the properties of the studied triacylglycerides (TAGs) are summarized in Supplementary Table 1 and those of the other lipophilic substances – in Supplementary Table 2. All substances were used as received. The TAG mixtures were prepared by mixing two or more substances in a molten liquid state, at the desired weight ratio.

The chemical structures and the specifications of the used surfactants are presented in Supplementary Table 3. Several very efficient and more complex in composition surfactant mixtures are denoted in the text as follows. Surfactant Mixture 1 (SM1) is a mixture of mono-acylglycerides $C_{16}MG:C_{18}MG:C_{18:1}MG$ in 55:40:5 weight ratio. Surfactant Mixture 2 (SM2) contains both mono- and di-acylglycerides in ratio MG:DG = 70:30 with saturated alkyl chains $C_{16}:C_{18}$ = 45:55. All surfactants were used as received. The aqueous phases were prepared with deionized water with resistivity > 18 MΩ·cm, purified by Elix 3 module (Millipore).

**Sample preparation**

The initial coarse oil-in-water emulsions, containing micrometre sized drops, were prepared either by rotor-stator homogenization with Ultra Turrax (IKA, Germany) or by membrane emulsification with (SPG) glass membranes when monodisperse drops were required.

**Microscopy observations**

For microscope observations, a specimen of the studied emulsion was placed inside a glass capillary with length of 50 mm and rectangular cross-section: width of 1 mm or 2 mm and height - 0.1 mm. This capillary was enclosed within a custom-made cooling chamber made of aluminium, with cut windows for optical observations. The chamber temperature was controlled by cryo-thermostat (JULABO CF30). The temperature in the chamber was measured with calibrated thermo-couple probe with an accuracy of ± 0.2°C. The thermo-probe was inserted in one of the orifices of the thermostating chamber and mounted at a position, where a capillary with the emulsion sample would be normally placed for microscope observations. In the neighbouring orifices the actual capillaries with the emulsion samples were placed. The correct measurement of the temperature was ensured by calibrating the thermo-couple with a precise mercury thermometer in the range of temperatures measured. In control experiments we heated the dispersions containing lipid micro-particles until their melting was observed. We always observed the melting process at temperatures very close, within ± 0.2°C, to the reported melting temperature of the bulk oil, $T_m$.

The optical observations were performed with AxioImager.M2m microscope (Zeiss, Germany). We used transmitted, cross-polarized white light, with included λ-compensator plate situated after the sample and before the analyser, at 45° with respect to both the analyser and the polarizer. Under these conditions, the liquid background and the fluid objects have typical magenta color, whereas the frozen birefringent areas appear brighter and may have intense colors.

**Drop size measurement**

The initial drop size was determination from microscopy images, captured in transmitted light with long-focus objectives of magnification ×10, ×20, ×50 or ×100. The mean size of the droplets obtained after one or several freeze-thaw cycles was determined by dynamic light scattering (DLS) on 4700C instrument (Malvern Instruments, U.K.), equipped with a solid state laser, operating at 514 nm. Multimodal software was used for analysis of the autocorrelation function of the scattered light. The results shown in Figure 2a are averaged from at least 3 measurements at scattering angles of 90°. The results shown in Figure 2b represent averaged results of several samples with the same chemical composition, but with different oil weight fractions – no significant effect of the oil volume fraction was observed up to ca. 20 vol. %. As an example, the blue curve in Figure 2b summarizes the results measured with all samples shown in Figure 2a.

**Experiments with bulk samples**

The drop size evolution in bulk emulsions was studied with 10 or 20 mL samples, placed in glass containers. These samples were first cooled for 3 h in a refrigerator at a temperature of 5°C to achieve complete oil drop freezing, followed by melting of the emulsion drops by placing the glass bottles in a thermostat and heating them at 1.5°C/min rate. The drop size distribution was determined by DLS measurements after each freeze-thaw cycles. These emulsion samples were inspected also by optical microscopy to check whether micrometer drops had remained undetected by the DLS measurements – no such drops were seen in any of the samples reported.

**DSC experiments**

The DSC experiments were performed on Discovery DSC 250 apparatus (TA Instruments, USA). The studied sample of individual TAG, mixture of TAGs, or TAG emulsion was weighted and placed into a DSC pan (Tzero pan, TA Instruments). Hermetic lid and Tzero sample press (Tzero hermetic lid, TA Instruments) were used to seal the DSC pan before measurements. The samples were cooled and heated with fixed rate, varied between 0.5 and 10 K/min. The DSC curves upon both cooling and heating were recorded. The integration of the DSC curves was performed using the build-in functions of the TRIOS data analysis software (TA Instruments).

**SAXS/WAXS experiments**

SAXS/WAXS measurements were performed using the Austrian SAXS beamline at Elettra Synchrotron, Trieste, Italy. The WAXS signal was recorded using Pilatus 100k detector and the SAXS signal – using Pilatus 1M detector. The working energy was 8 keV ($\lambda \approx 1.55$ Å). The samples were inserted into cylindrical borosilicate capillaries, which were placed into a thermostating chamber, similar to the one used for microscopy observations.

**Contact angle measurements**

Three-phase contact angle measurements were performed by placing a drop of the tested surfactant solution onto a solid lipid substrate and observing the profile of the formed sessile drop with DSA10 apparatus, Krüss, Germany. The lipid substrates were prepared by the following procedure: first, microscope slides were washed in alcoholic KOH and then hydrophobized by interacting with hexamethyldisilazane (HMDS). Afterwards, the melted lipid to be tested was placed on the hydrophobic slide and covered with a second hydrophobized slide to form an oil layer of homogeneous thickness. These layers were crystallized in a freezer when α-phase substrates were needed or, alternatively, they were left at room temperature for 2 days when β-substrates were prepared. The phase state of obtained substrates, α or β, was confirmed by DSC measurements.

**Acknowledgements**

The SAXS/WAXS measurements performed in Elettra Sincrotron, Italy (project # 20192017) were supported by project CALIPSOplus under Grant Agreement 730872 from the EU Framework Program for Research and Innovation Horizon 2020. The authors are especially grateful to Dr. Heinz Amenitsch (Elettra Sincrotrone) for the very valuable assistance during these measurements. The authors thank Mrs. Mila Temelska (Sofia University) for performing the contact angle measurements. This study was partially funded by the Proof-of-Concept grant ShipShape (#766656). The work of N.D. was partially funded by the Bulgarian Ministry of Education and Science, under the National Research Program "VIHREN", project ROTA-Active (№КП-06-ДВ-4/16.12.2019). The study falls under the umbrella of COST action CA17120 "Chemobrionics" funded by Horizon 2020 program.

**Author contributions:**

D.C. discovered the main phenomena described in this paper; D.C. and S.T. designed the study; D.G. and D.C. performed the experiments and summarized the results; D.C. analyzed the results and prepared the first draft of the manuscript; D.C., S.T. and N.D. clarified the mechanisms; N.D. edited and prepared the final draft of the manuscript; S.T. read critically the manuscript and suggested improvements; All authors participated in discussions and critically read the final manuscript.

# Supplementary materials

**Supplementary Table S1.** Properties of the triglycerides studied.

| Chemical name | Chemical formula | Abbreviation | Molecular weight, $M_w$, g/mol | Producer | Purity | Melting point, $T_m$, °C | Mass density, $\rho$, kg/m$^3$, [1] | |
|---|---|---|---|---|---|---|---|---|
| | | | | | | | liquid | β solid |
| 1,2,3-propanetriyl tridecanoate (tricaprin) | $C_{33}H_{62}O_6$ | $C_{10}TG$ | 554.8 | TCI Chemicals | > 98% | 31 - 35 | 891 (80°C) | - |
| 1,2,3-propanetriyl tridodecanoate (trilaurin) | $C_{39}H_{74}O_6$ | $C_{12}TG$ | 639.0 | Alfa Aesar | ≥ 99% | 45 - 47 | 880 (80°C) | 1057 |
| 1,2,3-propanetriyl tritetradecanoate (trimyristin) | $C_{45}H_{86}O_6$ | $C_{14}TG$ | 723.2 | Sigma Aldrich | ≥ 99% | 56 - 57 | 872 (80°C) | 1050 |
| | | | | TCI Chemicals | > 95% | 55 - 60 | | |
| 1,2,3-propanetriul trihexadecanoate (tripalmitin) | $C_{51}H_{98}O_6$ | $C_{16}TG$ | 807.3 | TCI Chemicals | > 85% | 64 - 68 | 866 (80°C) | 1047 |
| 1,2,3-propanetriyl trioctadecanoate (tristearin) | $C_{57}H_{110}O_6$ | $C_{18}TG$ | 891.5 | TCI Chemicals | > 80% | 66 - 74 | 863 (80°C) | 1043 |

The values of the melting temperatures are provided by the producers. The values of the mass densities are from Small, D. M. *The Physical Chemistry of Lipids. From Alkanes to Phospholipids* in Handbook of Lipid Research; Plenum: New York, 1986.

**Supplementary Table S2.** Properties of non-triglyceride hydrophobic substances studied.

| Chemical name | Chemical formula | Abbreviation | Molecular weight, $M_w$, g/mol | Producer | Purity |
|---|---|---|---|---|---|
| Hexadecane | $C_{16}H_{34}$ | $C_{16}$ | 226.4 | Sigma - Aldrich | ≥ 99% |
| α, α' - dilaurin | $C_{27}H_{52}O_5$ | $C_{12}DG$ | 456.7 | TCI Chemicals | > 96% |
| Dipalmitoylphosphatidylcholine | $C_{40}H_{80}NO_8P$ | DPPC | 734.0 | NOF Corporation | ≥ 99% |
| Progesterone | $C_{21}H_{30}O_2$ | - | 314.5 | TCI Chemicals | > 98% |
| Fenofibrate | $C_{20}H_{21}ClO_4$ | - | 360.8 | | ≥ 98% |

**Supplementary Table S3.** Properties and structural formulas of the surfactants used (all characteristics are taken from the descriptions provided by the surfactant producers).

| | Non-ionic surfactant (trade name/product No) | Number of C atoms, n | Number of EO groups, m | Producer | HLB | Structural formula |
|---|---|---|---|---|---|---|
| **Polyoxyethylene alkyl ethers $C_nEO_m$** | Brij 52 | 16 | 2 | Sigma - Aldrich | 5 | 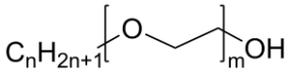 |
| | Brij 72 | 18 | 2 | | 4.9 | |
| | Brij 30 | 12 | 4 | | 9 | |
| | Brij C10 | 16 | 10 | | 12 | |
| | Brij S10 | 18 | 10 | | 12 | |
| | Brij 35 | 12 | 23 | | 16 | |
| | Brij 58 | 16 | 20 | | 15.7 | |
| | Brij S20 | 18 | 20 | | 15.3 | |
| **Polyoxyethylene Sorbitan monoalkylate $C_nSorbEO_m$** | Tween 20 | 12 | 20 | Sigma - Aldrich | 16.7 | 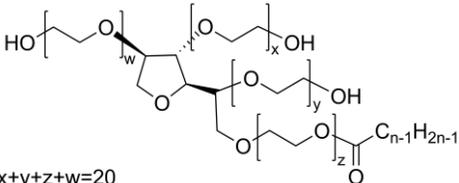 |
| | Tween 40 | 16 | 20 | | 15.5 | |
| | Tween 60 | 18 | 20 | | 14.9 | |
| | Tween 80 | 18 with double bond | 20 | | 15.0 | |
| **Polyethylene glycol monooleyl ethers $C_{18:1}EO_m$** | P0711 | 18 with double bond | 2 | TCI Chemicals | 4 | 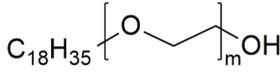 |
| | P0713 | | 7 | | 10.7 | |
| | P0714 | | 10 | | 12.4 | |
| | P0715 | | 20 | | 15 | |
| **Sorbitan monoalkylates $C_nSorb$** | Span 20 | 12 | - | Merck | 8.6 | 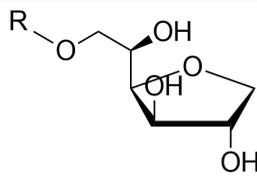 |
| | Span 40 | 16 | | TCI Chemicals | 6.7 | |
| | Span 60 | 18 | | Sigma-Aldrich | 4.7 | |
| | Span 80 | 18 with double bond | | Sigma-Aldrich | 4.3 | |
| **Polyoxyethylene octyl phenyl ether** | Triton X100 | octyl phenyl | 9.5 | Merck | 13.5 | 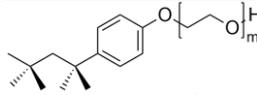 |
| **Monoglycerides $C_nMG$** | Monopalmitin | 16 | - | TCI Chemicals | - | 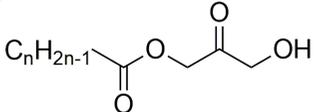 |
| | Monostearin | 18 | | | | |

| | Substances used as non-ionic surfactants (trade name/product No) | Number of C atoms, n | Number of EO groups, m | Producer | Purity | Structural formula |
|---|---|---|---|---|---|---|
| **Monoolein $C_{18:1}MG$** | 1-oleoyl-*rac*-glycerol | 18 with double bond | - | Sigma-Aldrich | > 40 % | |
| | Monoolein | | | TCI Chemicals | | |
| **Oleyl acetate** | A0934 | 18 with double bond | - | TCI Chemicals | > 60 % | |
| | **Ionic surfactants** | Number of C atoms, n | Hydrophilic head group | Producer | Purity | |
| **Anionic** | SDS | 12 | $SO_4^- Na^+$ | Sigma | 99 % | |
| **Cationic** | CTAB | 16 | $N^+(CH_3)_3\ Br^-$ | Sigma | > 99 % | |

**Supplementary Table S4. Cold-burst and double emulsion formation processes as observed with different $C_{14}TG$-surfactant combinations.** The surfactants were dissolved/dispersed in the aqueous phase in all experiments described in this table. The results are obtained in microscopy observations with emulsion samples placed in a glass capillary, the heating rate was varied between 0.5 and 2°C/min. See after the table for the legend used to describe process efficiency.

| Emulsifiers in the aqueous phase | | | Cold-burst process (for $d \leq 20$ μm) | Double emulsion formation |
|---|---|---|---|---|
| Water-soluble, 1.5 wt. % | Oil-soluble | | | |
| | Surfactant | c, wt. % | | |
| $C_{12}EO_{23}$ | $C_{12}EO_4$ | 0.5 | +++ | - |
| $C_{16}EO_{20}$ | $C_{18:1}EO_2$ | 1 | +++ | - |
| $C_{18}EO_{20}$ | - | 0 | ++ | - |
| | SM1 | 0.5 | +++ | - |
| | SM2 | 0.5 | +++ | - |
| | $C_{18}MG$ | 1 | +++ | - |
| | $C_{12}EO_4$ | 0.5 | +++ | - |
| | $C_{16}EO_2$ | 1 | +++ | - |
| | $C_{18}EO_2$ | 0.5 | +++ | - |
| | $C_{18:1}EO_2$ | 1 | +++ | - |
| | $C_{16}OH$ | 1 | +++ | - |
| | $C_{16}Sorb$ | 1 | ++ | - |
| $C_{16-18}EO_{50}$ | $C_{18:1}EO_2$ | 1 | +++ | - |
| $C_{12}SorbEO_{20}$ | - | 0 | + | - |
| | $C_{12}EO_4$ | 0.5 | +++ | - |
| | $C_{18:1}EO_2$ | 1 | +++ | - |
| | $C_{12}Sorb$ | 0.5 | +++ | - |
| | $C_{18:1}Sorb$ | 0.5 | - | ++ |
| | SM1 | 0.5 | +++ | - |
| $C_{16}SorbEO_{20}$ | $C_{16}Sorb$ | 0.5 | + | - |
| | $C_{18:1}EO_2$ | 1 | + | - |
| | SM1 | 0.5 | +++ | - |
| $C_{18}SorbEO_{20}$ | SM1 | 0.5 | + | - |
| $C_{18:1}SorbEO_{20}$ | $C_{18:1}Sorb$ | 0.5 | - | ++ |
| Triton X100 | $C_{18:1}EO_2$ | 1 | +++ | - |
| SDS | SM2 | 1 | ++ | - |
| SDS (0.5 wt. %) | - | 0 | - | ++ |
| | $C_{12}EO_4$ | 0.1 | + | - |
| | $C_{12}EO_4$ | 0.2 | ++ | - |
| | $C_{12}EO_4$ | 0.5 | +++ | - |
| | $C_{18:1}EO_2$ | 1 | +++ | - |
| | SM1 | 1 | +++ | - |
| $C_{16}TAB$ (0.5 wt. %) | - | 0 | + | - |
| | $C_{18:1}EO_2$ | 1 | ++ | - |
| | SM2 | 1 | +++ | - |

**Legend:** All descriptions are for the samples observed **after one** cooling and heating cycle.

|     | **Cold-burst process** | **Double emulsion formation** |
|-----|------------------------|-------------------------------|
| +++ | Very efficient cold-burst process, no micrometer drops are seen in the sample even after the 1st cooling-heating cycle. | Many small water droplets (< 2-3 μm in diameter) are trapped in each oily drop. |
| ++  | Most of the drops disintegrate efficiently to nanoparticles; small fraction of individual micrometer sized drops remains only. | More than 50 % of the oil drops contain several entrapped water droplets. |
| +   | Disintegration process is observed, but it is not very efficient to produce nanoparticles. The main fraction of the formed oil particles is with diameter of several μm. | Individual large in size water drops are trapped inside less than 50 % of the oil drops. |
| -   | The process is not observed. | |

**Supplementary Table S5.** Summary of the results, obtained with various triglyceride mixtures (1:1 in weight ratio), and their behavior upon one cooling-heating cycle, as observed in the microscopy experiments. The same symbols are used to denote the efficiency of the cold-burst process as in Supplementary Table S4.

| **Triglycerides in the mixture** | **Emulsifiers in the aqueous phase, wt. %** | | **Cold-burst process** |
|---|---|---|---|
| | **1.5% Water-soluble** | **Oil-soluble** | |
| $C_{12}TG + C_{14}TG$ | $C_{16}EO_{20}$ | 1% SM2 | +++ |
| | $C_{18}EO_{20}$ | 1% SM2 | +++ |
| | $C_{12}SorbEO_{20}$ | 0.5% SM1 | +++ |
| | | 0.5% $C_{18:1}EO_2$ | +++ |
| $C_{12}TG + C_{14}TG + C_{16}TG$ | $C_{18}EO_{20}$ | 0.5% $C_{12}EO_4$ | + |
| | | 0.5% $C_{18}MG$ | + |
| | | 0.5% SM1 | ++ |
| | | 1% SM2 | + |